# NoCFG: A Lightweight Approach for Sound Call Graph Approximation

Aharon Abadi
WhiteSource
Israel
aharon.abadi@whitesourcesoftware.com

Bar Makovitzki
The Open University of Israel
Israel
barmako@gmail.com

Ron Shemer
WhiteSource
Israel
ron.shemer@whitesourcesoftware.com

Shmuel Tyszberowicz
Afeka Academic College of Engineering
Israel
tyshbe@tau.ac.il

## ABSTRACT

Interprocedural analysis refers to gathering information about the entire program rather than for a single procedure only, as in intraprocedural analysis. Interprocedural analysis enables a more precise analysis; however, it is complicated due to the difficulty of constructing an accurate program call graph. Current algorithms for constructing sound and precise call graphs analyze complex program dependencies, therefore they might be difficult to scale. Their complexity stems from the kind of type-inference analysis they use, in particular the use of some variations of *points-to* analysis. To address this problem, we propose *NoCFG*, a new sound and scalable method for approximating a call graph that supports a wide variety of programming languages. A key property of *NoCFG* is that it works on a coarse abstraction of the program, discarding many of the programming language constructs. Due to the coarse program abstraction, extending it to support also other languages is easy. We provide a formal proof for the soundness of *NoCFG* and evaluations for real-world projects written in both *Python* and *C#*. The experimental results demonstrate a high precision rate of 90% (lower bound) and scalability through a security use-case over projects with up to 2 million lines of code.

## 1 INTRODUCTION

A program's call graph is a data structure that represents the runtime calling relationships between the procedures of a program [11, 42]. Call graphs are necessary prerequisites for most interprocedural analyses used in compilers, verification tools, and program understanding tools [6]. However, it is a difficult task to write algorithms that construct qualitative (i.e., sound and precise) call graphs. For example, implementations of state-of-the-art call graph construction algorithms for *Java* programs (e.g., WALA [52], SOOT [29]) are considered sound [13] though they are not fully sound due to different dynamic features of the language (e.g., reflection), and their unsoundness has been a topic for research on its own (e.g., [49], [38]). In this paper, we use the term *sound* in the same sense, i.e., we do not handle all the edge cases of the language [1]. Given the runtime types of all the variables of a program, one can use it to soundly infer the target of every method invocation site in the program to approximate a sound call graph of the program. Indeed, many sound approaches for call graph construction use *points-to analysis* [21, 44, 45, 48] to construct a call graph from its type inference data. While *points-to analysis* is sound and precise, it is expensive since, for example, it analyzes control-flow branching and different instances of the same types, limiting the scale of these algorithms. Name-based call graph construction techniques, i.e., considering method names alone, are imprecise and are not sound for functional languages [18] such as *Python*. In strongly typed languages such as *Java* or *C#*, call graphs can be soundly approximated by conservatively inferring the possible types of any receiver object from its static type [14]. However, these standard approaches sacrifice precision and fail to discover the call graph in a functional and dynamically-typed language [34] such as *Python*, due to the lack of static type declarations and dereference of function. As an example, consider the *Python* program listed in Fig. 1 (a) (adopted and adapted from [53]) and its call graph provided in Fig. 1 (b). The edge from *main* to *eat_bananas* (representing the invocation in line 22) can easily be discovered using the type of the variable *person* (defined in line 19); however, the type of the local variable *banana* in line 16 is dynamically set in line 21 and the previously mentioned methods will fail to infer its type.

***Our Approach.*** To construct a call graph approximation, we perform three steps. (1) Transform the program into an intermediate representation. (2) Infer the transformed program's types using an interprocedural analysis. (3) Build an approximated call graph based on the type inference data.

Step 1 removes all the program's control-flow constructs [5, 22] by replacing them with simple instructions that preserve their type-inferencing semantics. Therefore, the intermediate representation consists of only assignment, return, and method invocation statements. For example, the content of the *main* method (lines 19- 22 in Fig. 1 (a)) is transformed into the following:

person = Person()

---

[1] Also known as soundiness, see http://soundiness.org/.

```
1    class Banana:
2       def eat(self):
3          pass
4    class Carrot:
5       def eat(self):
6          pass
7    class Person:
8       def __init__(self):
9          self.no_bananas()
10      def no_bananas(self):
11         self.bananas = []
12      def add_banana(self, banana):
13         self.bananas.append(banana)
14      def eat_bananas(self):
15         for banana in self.bananas:
16            banana.eat()
17            self.no_bananas()
18   def main():
19      person = Person()
20      for a in range(1, 10):
21         person.add_banana(Banana())
22      person.eat_bananas()
```

(a)

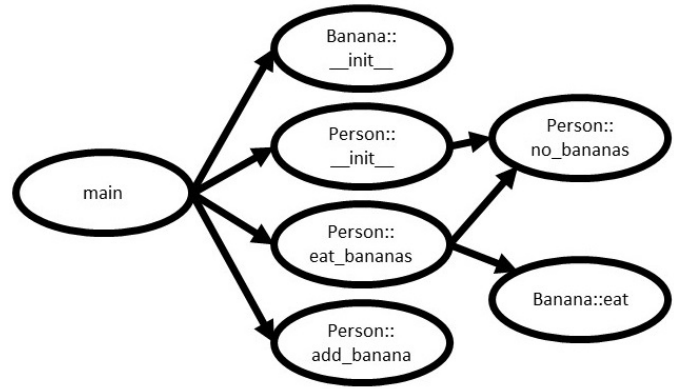

(b)

Figure 1: (a) A simple python program that represents dynamic type ; (b) The call graph of the program.

a = itemOf(**range**(1, 10))[2]
person.add_banana(Banana())
person.eat_bananas()

The *for* loop control (line 20) is replaced by an assignment that captures *Python*'s for loop semantics, keeping its type inferring side-effect consistent for the identifier *a* (i.e., that *a* is of type *Integer*), while the rest of code does not change. Notice that although this abstraction loses the control-flow data, it will be formally proved that it is sound and empirically shown that it has little to no effect over the precision.

The type inference is done by propagating facts defined by the intermediate representation statements: assignments extend the mapping of the assigned name by the types of the assigned expression, method invocations map the invoked method's parameters to the types of the invocation's arguments, and return statements define the types to which its method is evaluated to when it is invoked. Identifiers types (e.g., *person*, *Person*) can be derived from a prior assignment or from a name look-up. Member expressions types (e.g., *person.eat_bananas*) are derived in a similar manner, with a scope restriction to their object's types, i.e., considering only mappings and names relevant to those types. This means that for the above intermediate representation, the following happens: the first assignment (representing the assignment in line 19 from Fig. 1 (a)) adds a mapping from the name *person* to the types that the invocation of *Person()* is evaluated to, which in this case is solely the type *Person*, as the invocation target is the constructor of the class *Person*; *person.add_banana* is mapped to the method *Person:add_banana* since *person* is already mapped to *Person*; and the invocation *person.add_banana(Banana())* adds the type *Banana* to the mappings of the name *banana* from the *Person:add_banana* method.

---
[2]itemOf is defined over collection types and its semantics is to return a set of the types of its different items

Since the type inference analysis is already discovering reachable methods during the interprocedural analysis, the construction of the approximated call graph can occur simultaneously, placing an edge between two types (e.g. methods) once one invokes the other.

***Advantages of NoCFG.*** The presented method steps are similar to those of current state-of-the-art algorithms, as it uses type inference data to approximate the call graph of the program. However, it is fundamentally different by not considering any control-flow construct of the program, therefore allowing the analysis to work on a much simpler representation of the program and to remain sound.

Note that the process described above does not rely on any declared static type information and allows for methods to be referenced by variables through method types. This facilitates the analysis of a dynamic and functional language such as *Python*, and is applicable for many other languages such as *Java*, *C#*, *Ruby*, *JavaScript* and *C*. Furthermore, literal types (i.e., String, Integer, etc.) can be associated with their different literal values, therefore allowing the analysis to resolve common reflective invocation patterns, e.g., through *Python*'s *getattr* method. This is elaborated on in Section 3.1.

**Contribution.** We present a novel and simple approach for call graph approximation that is sound and uses a lightweight representation of program's code that discards many original program constructs, namely its control-dependency instructions. This lightweight representation is programming language independent, making it suitable for both static and dynamic programming languages, including programming features such as pointer types and first-class functions. We also show, using real-world programs, that this lightweight representation has a minor effect over the constructed call graph precision.

The rest of the paper is structured as follows: Section 2 provides the formal theory of the call graph approximation method, and Section 3 describes a detailed algorithm that implements the semantics. In Section 4 we describe various optimizations used in the implementation of the algorithm for the *Python* programming language and present our experimental evaluation. Section 5 discusses related work, and Section 6 concludes the paper.

## 2 THEORY

We now introduce the abstract syntax tree (AST [9]) representation and the operational semantics (inspired by [32]) we have applied. This semantics is based on a token-passing model which is very natural for proofs, as described in Abadi et al. [2]. Each type definition in a program is represented in the AST as a list of nodes, where each node represents a program statement. In *Python*, for example, such types include class-types, method-types and file-types. We denote by $N$ the set of statement nodes, where each node has one of the following disjoint types:

- *Assignment* ($N_A$), representing an assignment statement (e.g., $a = 3$), which consists of two expressions representing its left- and right-hand sides (*a* and *3*, respectively).
- *Method invocation* ($N_I$), representing a method invocation (e.g., *e.f(a)*) which has two parts: the target expression (*e.f*) and list of argument expressions (*[a]*).
- *Return* ($N_R$), representing a return statement (e.g., *return a*) and that has a single expression (*a*).

An expression ($E$) can be one of the following:

- *Name*, representing an identifier of the program. For example, *Banana*, *eat* and *self* are names in the program listed in Fig. 1(a).
- *Member access*, representing a field access and that consists of an object expression and a field name, e.g., *self.bananas*.
- *Literal*, representing a primitive program literal, e.g., the integer *1*, the string *"eat"*, etc.
- Method invocation, as described above.

Let $T$ be the set of all the types in a program, e.g., the program in Fig. 1 (a) contains the following types: *Banana, Banana:eat, Carrot, Carrot:eat, Person, Person:__init__, Person:no_bananas, Person:add_banana, eat_bananas, main*.

When handling control-flow constructs we keep their type-affecting parts (i.e., assignments, method invocations and return statements) and discard all the other instructions (e.g., *then*, *else*, etc). For example, the statement **if** *condition()* **then** *func1()* **else** *func2()* is translated into three nodes: *condition()*, *func1()*, and *func2()*. This transformation causes the statements of a method to be in a flat hierarchy and loses the order in which these statements are executed at runtime. To compensate for this information lost, we iterate through all the statements until a fixed point is reached. Later in the section we prove that this abstraction is sound.

We use a special form of *control-flow graph* (CFG [7]) which we call *simplified CFG* (SCFG).

DEFINITION 1 (SIMPLIFIED CONTROL-FLOW GRAPH). *An SCFG is a CFG that obeys the following properties: (1) it has a single entry node and a single exit node; (2) each node has a single predecessor and a single successor; (3) every node has one of the following types: method invocation, assignment, and return (each represents its respected statement type); (4) its nodes form a simple cycle.*

We associate every method definition with an SCFG. Figure 2 demonstrates the SCFG of the *main* method in line 18 from Fig. 1 (a). As mentioned above, every statement has a node representing it in the AST.

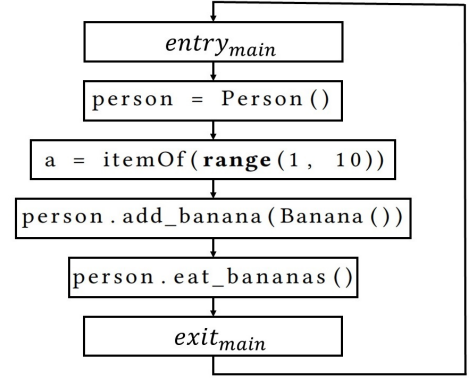

Figure 2: Simplified control-flow graph for the main method from Fig. 1 (a).

We now describe our call graph construction algorithm. We first address only its intraprocedural analysis details, and then we describe the interprocedural analysis details (in Section 2.2).

### 2.1 Intraprocedural Analysis

The formal operational semantics employed in this section is based on the plan calculus semantics [1, 39, 40]. For a given program $P$, let $\textbf{cflow}_P(n_1, n_2)$ denote the fact that its SCFG contains an edge from node $n_1$ to node $n_2$; let $\textbf{scope}_P: N \to T$ be a total function from nodes to type representations, such that for any node $n$ it returns the representation of the type containing $n$ (e.g., $\text{scope}_P(n)$ for any node $n$ in Fig. 2 returns the representation of the *main* method in line 18 of Fig. 1 (a)); let $\textbf{func}_P: T \times Name \to T$ be a partial function that for a given type and name returns its method representation (e.g., for the code example in Fig. 1 (a), $\text{func}_P(Banana, \text{'eat'})$ returns the type representation of the method *Banana:eat* in line 2)); let $\textbf{lhs}_P: S \times N_A \to V$ be a partial function that for a given state and an assignment node returns the variable on the left-hand side of the assignment based on that state, where an element of $V$ is a pair composed of type and name ($T \times Name$); let $\mathbf{rhs}_P: S \times N_A \to \mathcal{P}(T)$ be a partial function that for a given state and an assignment node returns the set of types that the expression on the right-hand side of the assignment can express using the mappings in the given state; let $\mathbf{control}_P: S \to N$ be a total function that for a given state $s$ returns the node the control-token is currently at; let $\mathbf{data}_P: S \times Var \to \mathcal{P}(T)$ be a total function that for any given state and variable returns a set of types this variable can reference in this state. Notice that for all functions, if a variable is not mapped to any type in a given state the result is an empty set.

In our semantics, each state consists of the current location in the SCFG of the program and an environment that provides types values for the program's variables. The type values are mappings from variables to sets of values that represent the types that every variable may reference at runtime. These type values are discovered throughout the analysis execution, as described below.

The model contains a special control token. At the beginning of the analysis execution, the token appears at the output edge of the entry node and all of the program's variables are mapped to an empty set of types. At each step of the execution model, the control token flows along the control edge (which is unique since there is only one successor for each node in an SCFG). When the control token reaches a node we say the node is activated, the types mappings are extended with the node's semantics (see Fig. 3), and the control token moves to the node's output edge. This node activation uses and updates the type mappings of the environment. For example, consider the assignment node that represents the statement $self.bananas = []$ in line 11 of Fig. 1 (a), the state's types mappings are used to evaluate the type of $self$ identifier of the left hand-side and extends this type with a new mapping with the name $bananas$ to the type(s) that the assignment's right-hand side is evaluated to (an array type), and the control token is be passed to the assignment node's successor.

A final state is reached when a fixed point is reached. A step is expressed by the function $step_P$, which takes a non-final state and returns the next state in the execution of $P$. The execution of $k$ steps of $P$ is denoted by the function $step_P^k$. In the rest of this paper, the $P$ subscript will be omitted in places where they are obvious from the context.

Formally, each state $s$ in the execution of a program $P$ consists of the following elements:

- The position of the current control token ($control_P$).
- The data environment mapping each variable to a set of types ($data_P$).

The semantics of $P$ is the partial function $[\![P]\!]$, which takes an initial state of $P$ and returns the state at which $P$ terminates. That is, $[\![P]\!](s) = step_P^k(s)$ if $s$ is an initial state and a fixed point is reached after $k$ steps.

$$\frac{n_1 \in N_A, \text{control}(s) = n_1, \text{cflow}_P(n_1, n_2),}{\text{lhs}_P(n_1) = v, \text{scope}_P(n_1) = f}$$
$$\frac{}{\text{data}(s', f, v) = \text{data}(s, f, v) \cup \text{rhs}_P(s, n_1)}$$
$$\text{control}(s') = n_2$$

**Figure 3: Assignment node semantics.**

LEMMA 1. *The intraprocedural analysis converges to a fixed-point.*

PROOF. Suppose that for some program $P$ the analysis does not converge to a fixed-point, i.e., the analysis execution of $P$ $(s_1, s_2, s_3, \ldots)$ is infinite. Thus, there exist a node $n$, a variable $v = (f, u)$, and an infinite subsequence of indices $i_1 < i_2 < i_3, \ldots$, such that:

$\forall j : \text{control}_P(s_{i_j}) = n \wedge \forall j \exists k < j : \text{data}_P(s_{i_j}, f, u) \neq \text{data}_P(s_{i_k}, f, u)$

Since the activation semantics of nodes only extends variables type mappings and $\forall s \in S : data_P(s, f, u) \in \mathcal{P}(T)$, it is also true that:

$\forall k \forall l < k : \text{data}_P(s_{i_l}, f, u) \subseteq \text{data}_P(s_{i_k}, f, u) \subseteq \mathcal{P}(T)$

Hence, $\forall j \exists k : \text{data}_P(s_{i_j}, f, u) \subset \text{data}_P(s_{i_k}, f, u) \subseteq \mathcal{P}(T)$, but since $\mathcal{P}(T)$ is finite, we derive a contradiction. □

## 2.2 Interprocedural Analysis

We now describe the interprocedural semantics. The semantics is similar to the call-string approach [3, 43]. It consists of the representation described in the intraprocedural case, that is, an SCFG representation for each method—where each SCFG has a single entry and a single exit to hold the incoming and outgoing interprocedural edges. Each call site is represented by two nodes, a method invocation node and its direct successor in the SCFG. The method invocation node, say $mi$, is connected to the entry node of the called method ($cm$) by an interprocedural *call edge*, and $mi$'s successor has an incoming interprocedural *return edge* from the exit node of $cm$. Note that method invocation nodes are connected to multiple edges: an intraprocedural edge to its successor node in its SCFG and interprocedural call edges to the called method(s) entry node. This implies that entry nodes may have multiple incoming call edges, and exit nodes may have multiple outgoing return edges. The representation of the program presented in Fig. 1 (a) is demonstrated in Fig. 4 and includes both intraprocedural and interprocedural edges.

We cannot assume that each method invocation uniquely identifies a target method for two reasons. (1) Object-oriented languages allow polymorphic dispatching, in which the code executed for a specific method invocation depends on the called object type at runtime. (2) Our type mappings allow for multiple types to be mapped to a single variable simultaneously and thus, for example, enables two methods to be mapped to the same variable. We support these situations by using a set of edges that connect each method call in the SCFG with the entries of all possible target methods. Using the type mappings obtained by the analysis we can *precisely* connect the method invocations and the possible targets, rather than over-approximating the set of targets by, e.g., the called method signature. The same technique can be used to handle method pointers in languages that offer them.

The semantics described in Section 2.1 must be extended. This is because nodes can have more than one outgoing/incoming edge in our graph representation, thus not all paths in it are valid paths, i.e., paths that preserve the semantics of method invocations. For example, if functions $f$ and $g$ call a function $h$, there would be two return edges from $h$ to both $f$ and $g$, allowing an invalid path as shown in Fig. 5. In addition, any valid path from the initial node may have unmatched calls (allowing, for example, for program termination by methods such as System.exit()), but cannot have unmatched returns. A *balanced path* is defined to be a valid path that has no unmatched calls or returns.

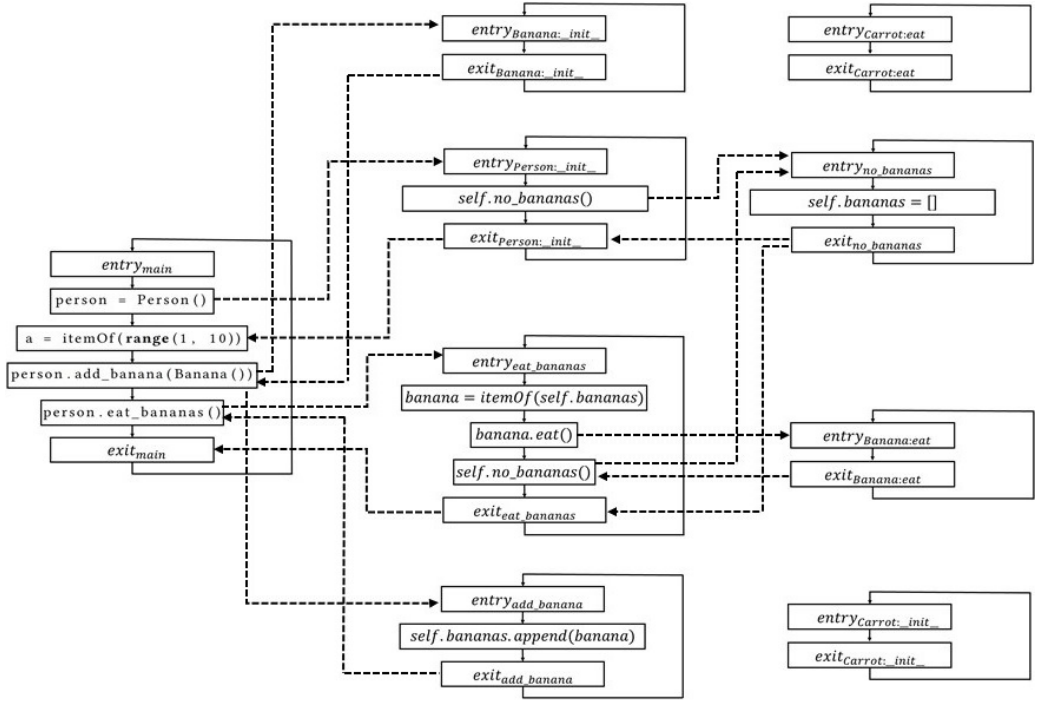

Figure 4: Full representation of the program from Fig. 1 (a). Interprocedural edges are represented by dashed arrows and intraprocedural edges are represented by solid arrows.

Formally, state $s$ in the execution of a program representation is extended with a stack function stack$_P(s)$ returning the stack at state $s$. When the token is placed before a call node, as a result of performing a single step in the execution, the control token is transferred to the entry of the called function and the corresponding successor node is pushed to the stack. When the token is placed before an exit node, as a result of preforming a single step in the execution the item at the top of the stack is popped and the control token is transferred to the resume node that is pointed to by the popped item. All other rules do not change the stack. The new rules are shown in Fig. 6 and Fig. 7. Note that these rules make sure the semantics covers only balanced paths.

In addition, we define the following functions: $\mathbf{expr}_P : N_R \to E$ be a total function from return nodes to expressions, such that for any node $n$ representing a return statement, it returns a representation of the expression in that statement, e.g., for the node representing **return** x it returns a representation of the expression x; $\mathbf{eval}_P : S \times E \to \mathcal{P}(T)$ be a total function from states and expressions to sets of types, such that for any state and expression, it returns the set of types that the evaluation of this expression can reference at runtime, e.g., for a simple identifier expression x and a state $s$ it returns the type mappings associated with the variable $(scopeOf(x), x)$ in $s$.

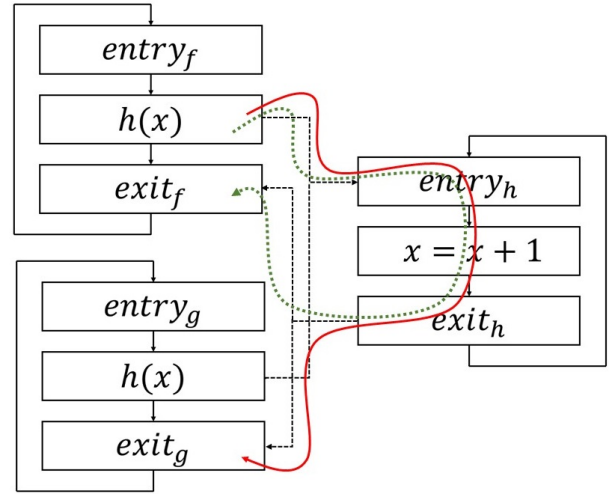

Figure 5: A small example of our graph representation that demonstrates a valid (dotted-green) path and an invalid path (solid red) going through interprocedural edges.

$$\frac{m \in N_I, \text{control}(s) = m, \text{cflow}(m, r)}{\text{control}(s') = m',}$$
$$\text{stack}(s') = \text{push}(\text{stack}(s), r)$$

Figure 6: Semantics for interprocedural method invocation. Method invocation arguments are bound to the target method parameters at state s', where m' is the entry node of a target method.

$$\frac{\text{control}(s) = c}{\text{control}(s') = \text{top}(\text{stack}(s)),}$$
$$\text{stack}(s') = \text{pop}(\text{stack}(s))$$

Figure 7: Semantics for interprocedural method exit, applied when we arrive at a local fixed-point. top(stack(s)) is the successor of the relevant method invocation node.

When the control token reaches a return statement node its expression is evaluated for its types and a special mapping is added to the type values of the state. This mapping uses a special variable for every method $m$, denoted $v_m = (m, \mu)$ (see Fig. 8). This special value is used when the control token passes through an interprocedural return edge to replace its calling method invocation with the types it holds.

$$\frac{n_1 \in N_R, \text{control}(s) = n_1, \text{cflow}_P(n_1, n_2),}{\text{expr}_P(n_1) = e, \text{scope}_P(n_1) = f}$$
$$\frac{}{\text{data}(s', f, \mu) = \text{data}(s, f, \mu) \cup \text{eval}_P(s, e)}$$
$$\text{control}(s') = n_2$$

Figure 8: Return node semantics.

DEFINITION 2 (INITIAL STATE OF SCFG). *Let $Q$ be a SCFG of program $P$. An* initial state *of $Q$ will be restricted to those that are equivalent to states of $P$ that can be reached by executions of $P$ without visiting any nodes of $Q$ before reaching those states.*

LEMMA 2. *Let $Q$ be a SCFG of program $P$. If $q_0$ and $q_1$ are distinct nodes in $Q$ such that $q_1$ is reachable from $q_0$ in $P$ through a path that does not contain any node in $Q$ except for the first and last nodes on the path, then $q_1$ is reachable from $q_0$ in $Q$.*

THEOREM 3 (CORRECTNESS OF EXECUTION WITH THE SEMANTICS). *Let $Q$ be a SCFG of a program $P$, $s_0$ an initial state of $P$, and $q_0$ the corresponding initial state of $Q$ (assuming one exists). If $P$ halts (i.e., reaches its exit node) when started at $p_0$, and a variable $x$ of a function $f$ gets a type $X$ during the execution, then $Q$ will also halt when started at $q_0$, and $X \subseteq \text{data}(s', f, x)$.*

PROOF SKETCH. Let $q_0, q_1, \ldots, q_n$ be the nodes from $Q$ that $P$ went through during its execution. (A node may occur more than once on this list, but we distinguish between these occurrences.) We prove by induction on the length of the sequence that the execution of $Q$, has the following properties: (1) it passes through $q_0, q_1, \ldots, q_n$; (2) if $s_k$ is the state of $P$ when it passes through $q_k$, then the state of $Q$ when it passes through $q_k$ over approximate the types calculated at $s_k$.

If the computation of $P$ does not go through $Q$ at all, the theorem is vacuously true. Otherwise, it is easy to show that the first node of $Q$ the computation enters is the one following $Q$'s entry, and the base step is true by the definition of an initial state of $Q$ (Def. 2).

Assume by induction that the claim is true for sequences of length up to $k$. If $k = n$, we are done. Otherwise, let $s_k$ be the state of $P$ when it reached $q_k$, and let $s'_k$ be the state of $Q$ when it reached $q_k$; by the induction hypothesis, $s'_k$ over approximate the types calculated at $s_k$.

Also, if the node to which $q_k$ belongs is an assignment or method call, it computes (or passes) the same values for its output data values in $P$ and in $Q$, and from the rules definitions it is easy to see that the theorem hold.

The next node of $Q$ that $P$ reaches is $q_{k+1}$; denote the corresponding state of $P$ by $s_{k+1}$. Note that the execution $P$ from $q_k$ to $q_{k+1}$ cannot change the state because it does not contain any assignment or method invocation. We need to show that $Q$ reaches $q_{k+1}$ with state required by the theorem. According to Lemma 2 there is a path from $q_k$ to $q_{k+1}$ and this implies that the paths in $Q$ looks like $q_0, \ldots, q', \ldots, q_{k+1}$.

Now consider the values at the input expression of the node to which $q_{k+1}$ belongs. Its source must be in $P$ (by the definition of a SCFG). If this values changes at $q'$ by the definitions of the rules they must be extended. All the rules just add types and never remove types. Hence the values at the input of the node in state $s'_{k+1}$ over approximate the values of all input variables calculated at $s_{k+1}$. From the definitions of the rules the value change at $q_{k+1}$ over approximate the value from $s_{k+1}$, as required to complete the induction step. □

## 3 ALGORITHM

In this section we describe the *NoCFG* algorithm that implements the semantics discussed in Section 2. The algorithm is inspired by the functional approach presented by Sharir and Pnueli [43] for computing the runtime type data (type inference) and the call graph for a given program. This algorithm is based on *Chaotic Iteration* [13], where the semilattice used reflects the information that the semantics is based on, namely the *data* function, i.e., mappings between program variables and sets of types. In the evaluation (Section 4) we describe the compact representation and some of the optimizations used to scale this algorithm to work with real world programs.

Let $S$ be the semilattice, $N$ be the set of nodes and $T$ be the set of types. The algorithm is worklist-driven and uses two main sub-functions: *partial-state* : $N \times S \rightarrow S$ and *summary* : $N \times S \rightarrow S$. For $n \in N$ and $s \in S$, *partial-state*$(n, s)$ returns a state $s'$ that is the product of propagating the facts from the entry node of *scope*$(n)$ with state $s$ to node $n$. For an entry node $n \in N$ and a state $s \in S$, *summary*$(n, s)$ returns a state that represents the execution of semantics over the statements of method *scope*$(n)$ with an initial state $s$. These two functions (mappings) are maintained throughout the analysis and are used to implement the functional approach of Sharir and Pnueli.

The algorithm (Algorithm 1) starts with an initial state that has no mappings at all (lines 1-2). Lines 3 to 39 implement a single chaotic iteration step, in which an item is drawn from the worklist, its relevant semantics is applied, and new items are added to the worklist. Each pair in the worklist holds the node to be handled and the initial state of the current method ($scope(n)$). This pair is in fact the relevant mapping key in *partial-state* to the current state. Lines 8 to 19 handle a method invocation node, where every possible target method is handled separately. Each discovered target method is connected with an interprocedural edge with its invocation (line 9) and either: (1) inserts a new workitem that holds the target's entry node and a state that has the target's parameters binded with the invocation's arguments type mappings to the worklist (lines 11 to 13), or (2) uses a previously summarized result of this invocation and continues with the invocation's successor (lines 15 to 17).

Lines 21 to 37 handle non-method invocation nodes. Each node's semantics is applied to the relevant state (via the *transform* function that is called in line 21). If the resulted state does not change (i.e., a local fixed-point is reached) it does the following: (1) local method variables mappings are removed from the resulted state (line 24); (2) *summary* mappings are extended with a mapping from the current method and the state used to start the method analysis to the resulted state (line 25); (3) a work item is pushed into the worklist to continue the analysis from where the current method has been invoked, using the *partial-state* mappings (lines 26 to 33). Lines 35 and 36 set the analysis for the successor of the current node, if the analysis did not arrive at a local fixed-point.

We will now describe the execution of the algorithm over the program in Fig. 1 (a). For simplicity, we assume that every variable in the program has a unique name in the program representation. The entry method for the algorithm is the entry of the method *main* defined in line 18. This node transform method does nothing, so the control passes to its successor node (assignment node representing *person = Person()*, line 19). The right-hand side method invocation is linked to its target, the constructor method of class *Person* in line 8. A new state with a binding between the parameter *self* to the type *Person* (as this is a constructor invocation) is created and the entry of this constructor is pushed into the worklist with the new state. The invocation in line 9 is then evaluated while the variable *self* of the method *__init__* in class *Person* is bound to the type *Person*. This binding enables the algorithm to determine that the target of the invocation is the method *no_bananas* of class *Person* in line 10. In a similar manner, the assignment node that represents line 11 can bind the type of the assignment right-hand side, which is the system type *array*, to the name *"bananas"* of type *Person*. When the algorithm returns from the evaluation of them constructor of *Person*, it also binds in place then return value of the constructor to the type of its class, i.e., *Person*, allowing the binding of the name *"person"* of type *main* to the type *Person*. *range(1, 10)* is a system method invocation which can be instantly be evaluated into an *array* of *Integers*, making the assignment representing the *for* header in line 20 bind the variable *"a"* of *main* to the type *Integer*. The invocation in line 21 uses the binding of *(main, "person")* to set its target to be the method *add_banana* of class *Person* (line 12), binding its *self* parameter with the type of *(main, "person")* and its *banana* parameter with the type evaluated from the invocation of *Banana()* which is the type *Banana* since it is a constructor

**Algorithm 1** Compute the types set for a node type C based on Sharir and Pnueli [43].

      callGraph($c : Type$):
1: worklist = ($e_{main}, \bot$)
2: Extend *partial-state by* partial-state($e_{main}, \bot$) =$\bot$
3: **while** worklist is not empty **do**
4:     ($n, s$) = worklist.getAndRemoveItem()
5:     let $f_n$ be the function containing $n$
6:     $s'$ = partial-state($n, s$)
7:     **if** $n$ is Method invocation **then**
8:         **for** $f$ target of $n$ in $s'$ **do**
9:             draw interprocedural edges between $n$ and $f$
10:           **if** summary($e_f, s'$) is not defined **then**
11:               Let $s''$ be $s'$ with appropriate parameter bindings mappings
12:               Extend *partial-state by* partial-state($e_f, s'$) = $s''$
13:               Add ($e_f, s'$) to worklist
14:           **else**
15:               Let $n'$ be the successor of $n$ in $f_n$
16:               Extend $s'$ with summary($e_f, s'$)
17:               Add ($n', s'$) to worklist
18:           **end if**
19:         **end for**
20:     **else**
21:         $s''$ = transform ($n, s'$)
22:         Let $n'$ be the successor of $n$ in $f_n$
23:         **if** partial-state($n', s$) = $s''$ **then**
24:             Remove $f_n$ variables mappings from state $s''$
25:             Extend *summary by* summary($e_{f_n}, s$) = $s''$
26:             **for** ($m, t$) in *partial-state* mapping **do**
27:                 **if** $m$ is the successor of a node $x$ calling to $f_n$ **then**
28:                     **if** partial-state($x, t$) = $s$ **then**
29:                         Extend *partial-state by* partial-state($m, t$) = $s''$
30:                         Add ($m, t$) to worklist
31:                   **end if**
32:               **end if**
33:             **end for**
34:         **else**
35:             Extend *partial-state by* partial-state($n', s$) = $s''$
36:             Add ($n', s$) to worklist
37:         **end if**
38:     **end if**
39: **end while**

invocation of the class *Banana*. The method invocation in line 13 then uses the existing mapping of *(Person, "bananas")* to conclude that the target of the invocation is the method *append* of the system type *array*, which allows for the invocation effect to be the extension of the items of this array by the types bound earlier to the parameter *(add_banana, "banana")*, i.e., the type *Banana*. When the algorithm starts handling the statements of *eat_bananas* of class *Person* in line 14, it first makes a mapping for the assignment representing the *for* loop header in line 15, since the mapping of *(Person, "bananas")* is to an *array* and it hold items of types *Banana* and *Integer*, the mapping for *(eat_bananas, "banana")* is extended with *Banana* and *Integer*. This last mapping allows the method invocation in line 16 to be associated with the method *eat* of type *Banana*. Table 1 shows the mappings of the final state.

Table 1: Type mappings in the final state of the analysis of the Python program in Fig. 1(a).

| Variable | | Types |
|---|---|---|
| Type | Name | |
| Banana:eat | self | {Banana} |
| Carrot:eat | self | $\phi$ |
| Person | bananas | {array} |
| Person:init | self | {Person} |
| Person:no_bananas | self | {Person} |
| Person:add_banana | self | {Person} |
| Person:add_banana | banana | {Banana} |
| Person:eat_bananas | self | {Person} |
| Person:eat_bananas | banana | {Banana, Integer} |
| main | person | {Person} |
| main | a | {Banana, Integer} |
| array | items | {Banana, Integer} |

### 3.1 Reflective API Support

The presented algorithm can be adapted to support common usage of reflection, by incorporating a simple variant of *string values analysis* [12] Consider the following *Python* code snippet that is taken from *Python*'s *ast* standard library [3], and uses a reflective API to implement a variation of the *visitor pattern* [4]:

**def** visit(self, node):
    method = 'visit_' + node.__class__.__name__
    visitor = **getattr**(self, method, self.generic_visit)
    **return** visitor(node)

We can approximate the types of calls like **getattr**(*self, method, self.generic_visit*) by keeping not only literal type(s) in the type mappings, but also their values. Assume the parameter *node* is mapped to *ast*-node types $Name$ and $If$. Then $node.\_\_class\_\_.\_\_name\_\_$ can be deduced into the type literals $String('Name')$ and $String('If')$. In a straightforward manner, the '$visit\_$' literal can be deduced into $String('visit\_')$. The assignment of the *method* variable can then infer a mapping between *method* to the type set {$String('visit\_')$, $String('Name')$, $String('If')$}. These literal values can then be used while approximating the targets of the reflective call to a set of methods reachable from the types of *self* with names matching any combination of them. For example, the *visitor* variable will be mapped to methods *visit_If*, *visit_Name* and *generic_visit* of any type that *self* is mapped to (assuming they exist).

## 4 EVALUATION

The presented algorithm can be used to approximate call graphs for programs written in a wide variety of programming languages. In this section we evaluate *NoCFG* for two popular programming languages: *Python* and *C#*.[5] *Python* is dynamically-typed and functional while *C#* is statically-typed, although each offer some support for the other's type-system: *Python* supports *type hints* [6] and *C#* offers a *dynamic* [7] type.

In this section we aim to demonstrate:

- The precision of *NoCFG*.
- The scalability of *NoCFG* through a real-world use case.
- *NoCFG* suits a wide variety of programming languages.

### 4.1 Implementation

We implemented the algorithmic core without the reflective API support described in Section 3.1. Dedicated front-ends were developed for *Python* and *C#*. Each such front-end implements an *Intermediate Representation* parsing mechanism and specific semantics for the *transform* and *parameters binding* operations (lines 21 and 11 of Algorithm 1).

The code is written in *Java* and all experiments were conducted with *Java HotSpot(TM)* 64-Bit Server VM (build 25.201-b09, mixed mode) running on a *ThinkPad T480* over *Windows* with heap size allocated limited to 8GB.

### 4.2 Python Evaluation

In Section 2 we provided a formal proof that *NoCFG* soundly approximates call graphs, yet soundness alone does not suffice for practical use cases. In fact, soundness may even not be difficult to achieve: a fully connected graph of all function calls is de facto sound. Therefore, a good balance between soundness and precision is required. To evaluate the precision of *NoCFG* we compared its results with a dataset of dynamically-generated call graphs. Table 2 lists 5 real-world open-source *Python* projects taken from *GitHub* for our benchmark.

Table 2: Information summary of *Python* benchmark projects

| Project | URL | LOC | Files |
|---|---|---|---|
| pytime | https://github.com/shinux/PyTime | 573 | 6 |
| algorithms | https://github.com/keon/algorithms | 12336 | 388 |
| purl | https://github.com/codeinthehole/purl | 839 | 11 |
| gspread | https://github.com/burnash/gspread | 1442 | 9 |
| wifi | https://github.com/rockymeza/wifi | 1031 | 15 |

To collect comparable pairs of dynamic and static call graphs for a given project we used its tests as entry-points for both dynamic and static analyses. Dynamic call graphs were collected by instrumentation the tests execution with *pycallgraph* [53].

Table 3 describes the results of the evaluation per benchmark project.

The results show high weighted averages of **89.66% precision** and **89.01% recall** for the benchmark projects. The fact that some feasible execution paths might not be covered by the tests suggests that the actual precision of *NoCFG* is even higher. Indeed, analyzing the over-approximated edges data shows that most of these edges are correct. For example, errors-raising execution flows that are not covered by a test are reflected in edges to an error constructor in the over-approximated edges set.

---
[3]https://docs.python.org/3/library/ast.html
[4]https://en.wikipedia.org/wiki/Visitor_pattern
[5]https://www.tiobe.com/tiobe-index/
[6]https://docs.python.org/3/library/typing.html
[7]https://docs.microsoft.com/en-us/dotnet/csharp/language-reference/builtin-types/reference-types#the-dynamic-type
[8]reflects a lower bound for the actual precision

Table 3: Summary of call graphs comparison between *NoCFG* and dynamic analyses. Both analyses used the tests as entry-points.

| Project | Matched edges | Over approx. edges | Missed edges | Precision [8] | Recall | Entry points | Time |
|---|---|---|---|---|---|---|---|
| pytime | 60 | 12 | 0 | 83.33% | 100.00% | 6 | 16s |
| algorithms | 810 | 94 | 109 | 89.60% | 88.14% | 392 | 13m 20s |
| purl | 232 | 20 | 9 | 92.06% | 96.27% | 102 | 1m 5s |
| gspread | 189 | 28 | 34 | 87.10% | 84.75% | 41 | 1m 39s |
| wifi | 70 | 3 | 16 | 95.89% | 81.40% | 48 | 19s |

Due to the soundness of the algorithm, one would expect the recall being perfect (i.e., 100%) for all of the projects but the lack reflective API support impairs the recall. This is confirmed by going over missed edges set.

### 4.3 C# Evaluation

For our C# evaluation we define a project as vulnerable if its program invokes a method known to be vulnerable. We are interested in deciding for real-world projects whether they are vulnerable due to a known vulnerable method located in a package referenced by the project. Using call graph terminology - we want to find out if a vulnerable method is reachable from the project code.

For this purpose, we distinguished application code (defined in the project) from extension code (defined in external packages). Then we manually over-approximated the control flow of the program and found a sound set of application entry points that may lead to the vulnerable method.

In this evaluation we examine open-source projects hosted on *GitHub* that are reported to reference packages that are affected by *CVE-2019-0820* [9]. The analysis results are presented in Table 4. For vulnerable projects we manually verified the method traces, while for non-vulnerable projects we relied on the projects' tests to verify that the vulnerable method is not invoked. The results show that *NoCFG* can be used to build large scale sound call graphs with no false positives in the examined projects.

Overall, we can see that only 4 out of the 11 projects that reference a vulnerable package are effectively vulnerable. Therefore, with this knowledge a maintainer of a non-vulnerable project may prioritize remediation for other CVEs that are effective, regardless of the vulnerability severity, which is the common prioritization metric.

In Fig. 9 we present a method call trace detected by the analysis in the project *EPPlus*. We manually followed the invocations and verified that the trace represents an actual program flow. The trace starts at the application method *GetValuesFromString* and reaches the vulnerable method *Go* that is defined in *RegexInterpreter*. The last edge in the path that reflects the presented call trace is from *Scan*, defined in the abstract class *RegexRunner* to *Go* which is an abstract method overridden in *RegexInterpreter*, a direct subclass of *RegexRunner*

```
System.Text.RegularExpressions.RegexInterpreter:Go
System.Text.RegularExpressions.RegexRunner:Scan
System.Text.RegularExpressions.Regex:Run
System.Text.RegularExpressions.Regex:Match
System.Text.RegularExpressions.RegexReplacement:Replace
System.Text.RegularExpressions.Regex:Replace
OfficeOpenXml.FormulaParsing.Excel.Functions.DateTime.TimeStringParser:GetValuesFromString
```

---

[9] https://msrc.microsoft.com/update-guide/vulnerability/CVE-2019-0820

Figure 9: A vulnerable call graph path detected by *NoCFG* in a C# project. The path starts at user-written code (*GetValuesFromString*) and reaches a vulnerable method (*Go*) in one of the project's external libraries (reported by *CVE-2019-0820*)

## 5 RELATED WORK

Propagation based program analysis for call graph approximation are well known and have been studied for many years. When designing a call graph approximation method there is always a tradeoff between soundness, precision and scalability. Simple approaches such as a name-based resolution, e.g., *Reachability Analysis* [47], *Class Hierarchy Analysis* (*CHA* [14]) and *Rapid Type Analysis* (*RTA* [10]), are imprecise and are not sound for dynamic and functional languages. Shivers proposed *Control Flow Analysis* (*CFA* [44, 45]) as a more advanced method for sound call graph approximation, which uses *points-to* analysis to infer the runtime types of every object expression in a program. Many different and popular methods were based on *CFA* for dynamically object-oriented languages [4, 35, 36], functional languages [23, 44] and statically object-oriented languages (e.g., *Java*) [15, 24, 50]. Tip and Palsberg [51] explored the design space between *RTA* and *CFA*) and proposed a few *unsound* approaches that balance between the accuracy and the computational price of the analysis.

*Type Analysis for JavaScript* (*TAJS*, [27]) is a sound *JavaScript* static program analysis infrastructure that infers type information that has been used to build tools for refactoring of *JavaScript* [16, 17] and enables a technique of statically resolving *eval* constructs [26]. However, *TAJS* combines heap and value (type) analysis that works on the CFG of the program, which results in a complicated lattice structure and transfer functions, therefore tempering its ability to scale. Indeed, the authors claim that *TAJS* "is targeted at handwritten programs consisting of a few thousand lines of code".

*Phantm* [28] is a *PHP 5* static analyzer for type mismatch based on data-flow analysis. *Phantm* is flow-sensitive, hence utilizing the control-flow data of the program and omits updates of associative arrays and objects with statically unknown values and aliasing, which makes it unsound.

Sridharan et al. [46] presented a static flow-insensitive points-to analysis for *JavaScript* that improves the precision and performance of the analysis for a reoccurring coding patterns in popular *JavaScript* frameworks.

Jang and Choe [25] presented a flow-insensitive points-to analysis that is based the classical points-to algorithm by Andersen [8].

Table 4: Summary of C# projects evaluation. LOC counts the lines of code in application and its decompiled external dependencies. The Vulnerable column reports *NoCFG*'s findings about whether the vulnerable method of *CVE-2019-0820* is reachable from an application node (i.e., method)

| Project | URL | LOC | Vulnerable | Time |
| --- | --- | --- | --- | --- |
| App.Metrics.Core | https://github.com/AppMetrics/AppMetrics | 182,556 | ✓ | 4m 33s |
| CoreRPC.AspNetCore | https://github.com/kekekeks/CoreRPC | 377,403 | ✗ | 20m 2s |
| EasyCaching.Bus.RabbitMQ | https://github.com/dotnetcore/EasyCaching | 83,040 | ✗ | 8m |
| EPPlus | https://github.com/JanKallman/EPPlus | 413,377 | ✓ | 14m 31s |
| FastReport.Data.RavenDB | https://github.com/FastReports/FastReport | 752,664 | ✗ | 3m 38s |
| Nucleus.Web.Vue | https://github.com/alirizaadiyahsi/Nucleus | 2,017,040 | ✗ | 1m 6s |
| Ooui.AspNetCore | https://github.com/praeclarum/Ooui | 839,959 | ✗ | 5m 28s |
| RawRabbit.Enrichers.MessagePack | https://github.com/pardahlman/RawRabbit | 104,003 | ✗ | 41s |
| SmartCode.ETL.PostgreSql | https://github.com/dotnetcore/SmartCode | 1,214,542 | ✗ | 1m 4s |
| SmartCode.Generator | https://github.com/dotnetcore/SmartCode | 1,238,470 | ✓ | 2m 4s |
| YoutubeExplode | https://github.com/Tyrrrz/YoutubeExplode | 197,448 | ✓ | 13m 41s |

It models variables, arrays and objects as associative arrays. Limitations of their work include that it precisely models only assignments to constant indices while using an unknown field for other cases.

*Code2Graph* [20] and *Pyan* [54] are two static call graph tools frequently mentioned in the developer community [31]. Both use a simplistic analysis that is not sound. Namely, these analyses do not handle many functional programming patterns which are popular in *Python*.

A Recent work on call graph construction for *C#* by Garbervetsky et al. [19], which focus on offering a generic distributed call graph construction framework based on existing type inference approaches. The implementation of this work relies on *RTA* for type inference which implies the call graphs it generates are not sound, while the use case described in our *C#* evaluation assumes the use of a sound call graph for the analysis.

Other related works from recent years describe *unsound* approaches dedicated for a specific usecase. Feldthaus et al. [18] proposes two AST-based approaches, much like ours, that are field-based and only track function objects. Reif et al. [37] describes two *CHA* variants, each is *sound* under an assumption that may be applicable in certain contexts.

## 6 DISCUSSION AND CONCLUSIONS

We have presented *NoCFG*, a novel and sound approach for call graph construction. This approach can be further optimized by preserving some control-flow data through an adaptation in the code representation (SCFG). Rather than connecting the exit and the entry nodes of the SCFG, we can modify the representation to have intraprocedural "loop" edges over loops constructs; see Fig. 10 for an example. This optimization will improve both the precision and the convergence time of the algorithm.

Furthermore, the adaptation for handling reflective API calls (mentioned in Section 3) can be explored with different string similarity metrics, e.g., Levenshtein's distance [30], and evaluated against different scenarios.

It would be interesting to extend our *C* proof-of-concept implementation of the analysis to a fully-fledged implementation with the optimizations described above, and to conduct a comparative study against different sound call graph construction approaches.

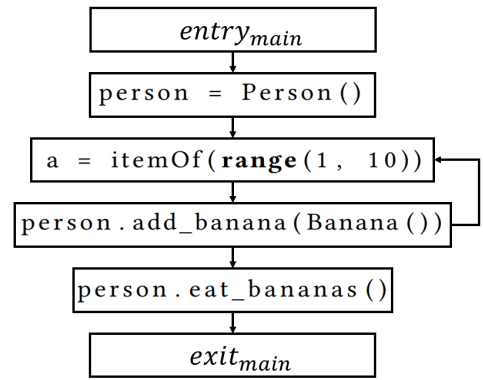

Figure 10: Adapted simplified control-flow graph for the main method from Fig. 1 (a).

Another interesting research subject is to extend *NoCFG* to distinguish between direct manipulation of types and manipulation of their type instances (i.e., instance-objects manipulation should not affect its type(s) representation and therefore any other instance of that type) and explore how it affects precision and scalability.

Our approach works on a simple representation of the code, thus removing the need to construct expensive code representations that require different control-flow or data-flow analyses.

We have implemented and evaluated *NoCFG* for *Python* and *C#*. The experimental results show the analysis is precise and scalable for real-world projects, despite its simplified code representation. These characteristics make *NoCFG* suitable for different code-analysis applications that require a whole application interprocedural application call graph for a wide variety of programming languages.

11